RAO-SS: A Prototype of Run-time Auto-tuning Facility for Sparse Direct Solvers[*]

Takahiro Katagiri[†,††], Yoshinori Ishii[†], Hiroki Honda[†]

† Graduate School of Information Systems,
The University of Electro-Communications
†† PRESTO, Japan Science and Technology Agency

1-5-1 Choufu-Gaoka, Choufu-shi, Tokyo 182-8585, JAPAN
E-mail: katagiri@is.uec.ac.jp[†]
Tel: +81-424-43-5642
FAX: +81-424-43-5644

**Abstract**
In this paper, a run-time auto-tuning method for performance parameters according to input matrices is proposed. RAO-SS (Run-time Auto-tuning Optimizer for Sparse Solvers), which is a prototype of auto-tuning software using the proposed method, is also evaluated. The RAO-SS is implemented with the Autopilot, which is middle-ware to support run-time auto-tuning with fuzzy logic function. The target numerical library is the SuperLU, which is a sparse direct solver for linear equations. The result indicated that: (1) the speedup factors of 1.2 for average and 3.6 for maximum to default executions were obtained; (2) the software overhead of the Autopilot can be ignored in RAO-SS.

---

## 1. Introduction

Recently, the demand for high performance processing to matrix computations is getting increased in computational science field. To obtain high performance in pieces of software in the filed, it is a reasonable way to adapt them to high performance numerical libraries.

We can classify the numerical libraries into two categories. That is, dense solvers and sparse solvers.

For remarkable characteristics for the sparse solvers, sparse solvers have many choices to optimize their parameters to establish high performance according to the location of non-zero elements for input matrices. There is no room to be considered for the optimization in dense solvers for the location of input matrices. The settings for the parameters in sparse solvers are now getting more and more complex. For library user's point of view, such settings are time-consuming and waste-of-time matters. In addition, if the user specifies a wrong parameter, the performance may be excessively low, or the computer may be break-down in execute-time because of fault memory allocation. This situation hardly causes dense solvers, since the users can know the size of matrix in advance.

Not in connection with dense and sparse solvers, default parameters are set in a constant value to avoid omitting the definition for the parameters. For this reason, we cannot obtain better performance in several kinds of computer environments. Moreover, the characteristics of input matrix are not sufficiently used to optimize the performance.

With respect to these parameter setting matters, there are many studies for automatic parameter setting and library construction methodology for numerical libraries. Especially, auto-tuning techniques to set appropriate parameters have been aggressively studied.

There are two kinds of auto-tuning timing for numerical library[8]. The first one is known as install-time auto-tuning, which is performed when the target software is installed. The second one is known as run-time auto-tuning, which is performed when the target software is running. In this paper, taking into account of the nature of sparse solvers, the run-time auto-tuning is focused.

For conventional software for the run-time auto-tuning, the Autopilot[1] and the ActiveHarmony[7] were developed. The I-LIB, which is a hybrid type of auto-tuning software between install-time and run-time auto-tuning, were proposed. The FIBER, which was proposed by T.Katagiri et.al, had introduced a new auto-tuning timing, named Before Execute-time Optimization (BEO)[5], in addition of the above two kinds of timings. The BEO is performed when the user fixes special parameters that are called Basic Parameter in their framework.

The aim of this paper is to propose a method to obtain high performance by setting appropriate performance parameters at run-time. The software implemented the proposed method is called RAO-SS (Run-time Auto-tuning Optimizer for Sparse Solvers)[6]. The RAO-SS has been implemented with the Autopilot to support auto-tuning facility at run-time. The goal of this paper is to propose the functionality and to evaluate the performance efficiency for RAO-SS.

This paper is organized as follows. Section 2 explains the Autopilot. Section 3 explains the SuperLU, which is the target application of this study. Section 4 proposes a facility of run-time auto-tuning and RAO-SS. Section 4 also describes the implementation details for RAO-SS. Section 5 is for performance evaluation of RAO-SS. Finally, we summarize some observations in this study.

## 2. Autopilot

### 2.1 Overview

The Autopilot is middle-ware to define complex computer management requirements. For example, to maximize performance of distributed parallel machines, some collisional requests are usually needed, e.g. minimizing latency and maximizing bandwidth. To describe such requests, the Autopilot introduces the function of fuzzy logic as a principal function[1]. It is a remarkable characteristic of RAO-SS to use the fuzzy logic function in the Autopilot to support auto-tuning process. Since the Autopilot is invoked as a daemon process in the system, we can adapt the auto-tuning facility to the GRID portal[3] or heterogeneous computer environments connected a network.

The Autopilot provides several instruction sets, such as Performance Sensor, Decision Procedure, and Policy Actuator, to perform adaptable control and resource management. The Autopilot also supports the functions to send and receive data. The Sensor is a function to send data, and the Actuator is a function to receive data. The Decision Procedure, which is an important function to select performance parameters based on received data from Sensor, acts an important roll for the implementation in RAO-SS. The Fuzzy Logic, which is a distinctive function of the Autopilot, also can be used to implement the auto-tuning facility. The auto-tuning facility is an unaware process for the library users---the process determines appropriate parameter sets based on the characteristics of input matrices.

### 2.2 Decision Procedure

The Decision Procedure in the Autopilot can select parameters based on user-defined policy, which can be defined with the behavior of applications and systems. The procedure gives us adaptable control in particular application on parallel environments, or widely distributed systems according to the system resource management policy.

Fig.1 shows an example of Decision Procedure with Fuzzy Logic.

First, information data for the input matrix is received from the user application or the system via Sensor. Second, the data is sent to Fuzzy Logic Rule Base from Sensor, and then, the parameter is determined based on the user-defined policy in Decision Procedure. Finally, the determined parameters in Decision Procedure are sent to the user application or the system via Actuators.

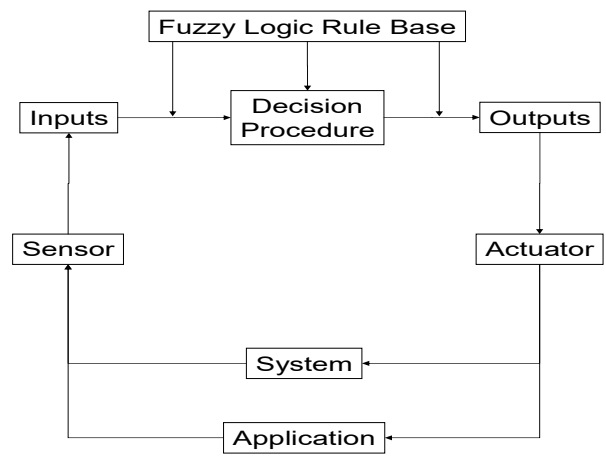

Fig.1 The Autopilot Fuzzy Logic Decision Procedure

## 3. SuperLU

### 3.1 Overview

The SuperLU[2] is a high performance numerical library to solve linear equations with the coefficients of large, sparse, and non-symmetric with the direct method of LU decomposition. The SuperLU consists of the forward elimination routine and the backward elimination routine with partial pivoting. For the released package, there are three kinds of types, which are for sequential machines, for shard memory machines, and for distributed memory machines.

The sequential SuperLU package versioned 3.0 is used in this paper.

3.2 Ordering Parameters

The followings show the performance parameters used in the SuperLU.

(1) Ordering Parameters
 (1.1) Approximate Minimum Degree Column Ordering
   (The default parameter.)
   Abbreviation: **COLAMD**
 (1.2) Natural Ordering
   Abbreviation: **NATURAL**
 (1.3) Minimum Degree Ordering on Structure of $A^T+A$
   Abbreviation: **AT+A**
 (1.4) Minimum Degree Ordering on Structure of $A^T * A$
   Abbreviation: **AT×A**
(2) Panel Size
(3) Realization Parameter
(4) The Minimum Number of Super-Node
(5) The Minimum Row Length for 2-D Blocking
(6) The Minimum Column Length for 2-D Blocking

We focus on the parameter of ordering. For the ordering algorithm, many methods were known. For example of trianglization method[10], the Markowitz method and the Tewarson method were proposed. For reduced band length method, the RCM method and the Minimum Ordering method were known. For blocking method, the Stewart method and the Dissection Division method were publised[4].

We focus on the auto-tuning method for the ordering parameter at run-time in RAO-SS, since the best ordering parameter extremely depends on the location of zero elements for input matrix.

4. RAO-SS: A Prototype of Run-time Auto-tuning Facility for sparse direct solvers

4.1 Overview

RAO-SS is a prototype of the run-time auto-tuning facility to establish high performance for the target library by specifying appropriate performance parameters automatically. Fuzzy logic is used for the Decision Process in the Autopilot. The target numerical library is the SuperLU, but not limited to it in the nature of our strategy.

4.2 Design Policy

RAO-SS is based on the following design policy:

(1) To obtain the characteristics of input matrices at run-time, the measure of *matrix density* is used.
(2) To implement the process of parameter decision easily, the Decision Process in the Autopilot is used.
(3) To add the facility of RAO-SS without re-compiling to legacy codes or object codes, the Autopilot is used as middle-ware.
(4) The aim is to prevent excessively speed-down with default parameter for the ordering for each jobs. It is not aimed to improve the total system throughput.

The design policies for (2) and (3) are acceptable if the principal function of the Autopilot is used.

Fig.2 shows the structure of the flow for RAO-SS.

First, the matrix density for input matrix is calculated. The matrix density is defined as the following Formula (1).

Matrix Density :=
The number of non-zero elements / (N*N)
  *100  [%]  …(1)

The N is matrix dimension. The calculated matrix density is sent to Sensor, which is stationed as input daemon for the system. The stationed process determines the appropriate parameters in the Decision Process on the Autopilot based on the received matrix density. The determined parameter, in this case, the ordering parameter, is sent to the output daemon. Then the output daemon sends the data to the user process via the SuperLU wrapper interface. Since the SuperLU is the user process in this case, the received ordering parameter can be directly set in the user application of the SuperLU.

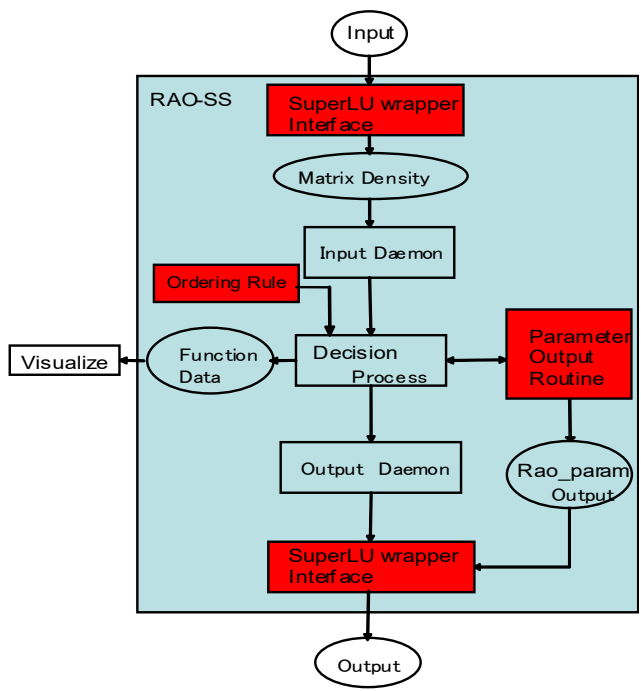

Fig.2 The system configuration in RAO-SS. The shade boxes imply the modification from the original routines in the Autopilot.

4.3 The Strategy for Parameter Selection

The matrix density is calculated by Formula (1). Then, the ordering parameter of the SuperLU is automatically determined with the matrix density in RAO-SS. Generally speaking, the best ordering parameter depends on the location of non-zero elements. If the user chooses the parameter groundlessly, the system may be down, since the system tries to allocate huge memory space without consideration of the location for non-zero elements. In this paper, we do not treat the functionality---that is, recovery facility for the system if the auto-tuning facility specifies a wrong parameter set.

Fig.3 illustrates the strategy for ordering parameter with the matrix density.

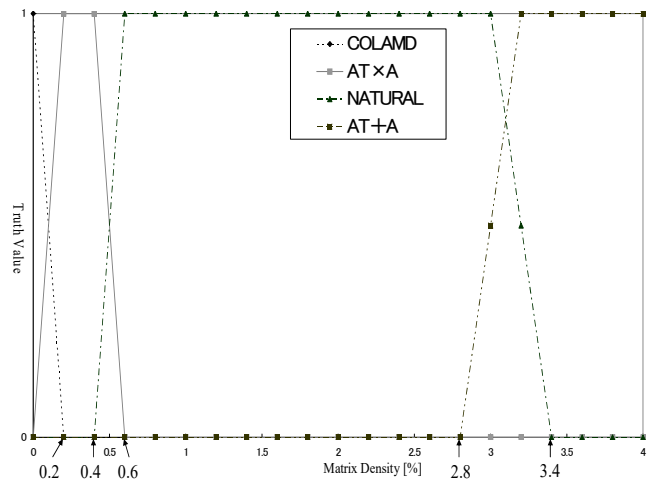

Fig.3 The strategy for parameter selection based on the matrix density in RAO-SS.

Fig.3 shows that the ordering parameter that is near to the value of 1.0 in Y-axis is chosen according the matrix density in X-axis. If there are two parameters in the same matrix density, it means that both of them can be selected. But only one parameter should be selected at run-time. The decision in this case is performed by the Fuzzy Logic Function in the Autopilot. For example, in the case of 0.5 in the matrix density, the considered ordering methods are $AT \times T$ and NATURAL. Only one method, however, should be determined at run-time in Fuzzy Logic Function with respect to the strategy of fuzzy logic in the Autopilot.

### 4.4 Implementation Details

The assumption of implementation for RAO-SS is to install the SuperLU and the Autopilot. As we explain in Section 3, the sequential version of SuperLU is used. To install the Autopilot, the Globus[11)] should be installed before.

We have implemented the strategy for parameter selection shown in Fig.3. The strategy in Fig.3 was determined empirically with the test matrices described in later section. Our method, hence, is sorted as an empirical optimization method for auto-tuning.

We modified the original source codes in the Autopilot to implement the processes of RAO-SS (See Fig.2.)

## 5. Performance Evaluation

### 5.1 Computer Environment

In this section, we evaluate the performance of RAO-SS and the SuperLU (hereafter the SuperLU denotes SLU.) We used a PC in this performance evaluation. The spec of the PC is shown as follows: OS:RedHat Linux7.8; CPU : Intel Pentium4 (1.8GHz); Memory : 512MB; Compiler: gcc ver.2.96;

The version of the Autopilot is 2.4.

The 32 kinds of matrices were used for the test matrices. These matrices are non-symmetric matrices, and come from the University of Florida Sparse Matrix Collection[12)]. The matrix densities we treated in this paper were from 0.004% to 5.5%.

### 5.2 A Survey for Optimal Parameter

First, we surveyed the best ordering parameter (hereafter we denote OP) in the Pentium4 for SLU. Fig.4 and Fig.5 show the speedup ratios between the best OP and default OP.

From the result of Fig.4 and Fig.5, the maximal speedup factor was about 4151 times. The matrix was *circuit4*. Almost 20 times speedup was also obtained. The matrix was *Mark3jac080sc*. The other matrix should be mentioned here was *circut3*. In this matrix, about 6.5 times speedup was obtained. These speedups are crucial.

It also showed that the default OP of SLU was excessively inefficient for some input matrices.

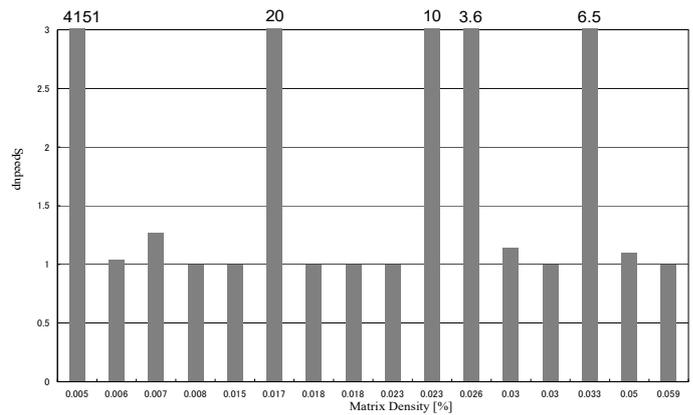

Fig.4 Speedups in the best OP to default OP. (Matrix Density: 0.005%--0.059%)

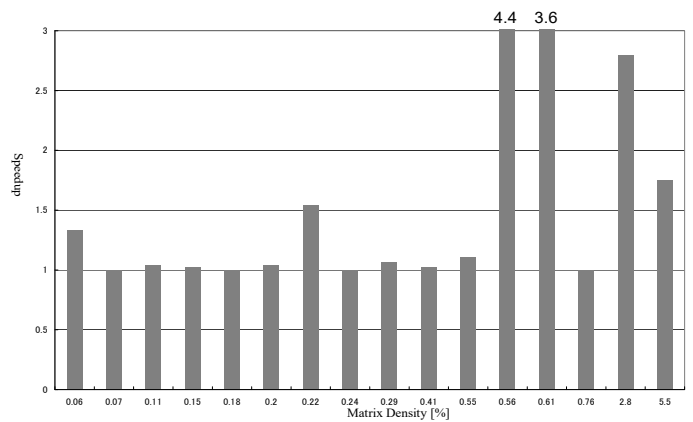

Fig.5 Speedups in the best OP to default OP. (Matrix Density: 0.060%--5.5%)

### 5.3 Performance of the SuperLU Implemented The Proposed Strategy

In this section, we evaluate the SLU implemented with the proposed strategy for ordering parameter selection with the matrix density. We denote the new SLU as SLU' hereafter.

By using the empirical strategy for the proposed method, the best OP is not always selected. The aim of this evaluation is to evaluate the performance with automatically selected OP compared to the performance with the best OP.

Fig.6 and Fig.7 show the speedup factors to executions with default OP. The matrices in which the speedups exceeded 1.0 have a potential for the improvement of auto-tuning with RAO-SS.

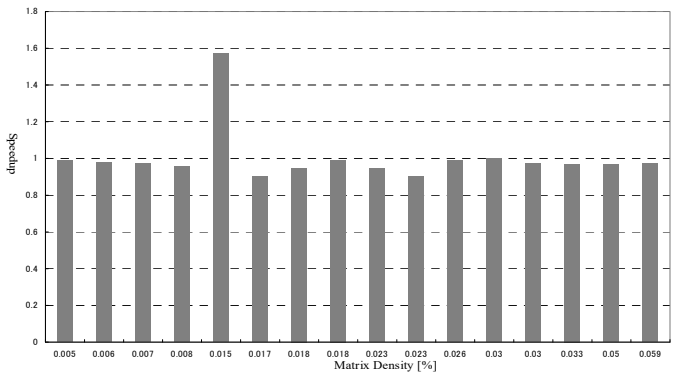

Fig.6 Speedups in SLU' to original SLU.
(Matrix Density: 0.005%--0.059%)

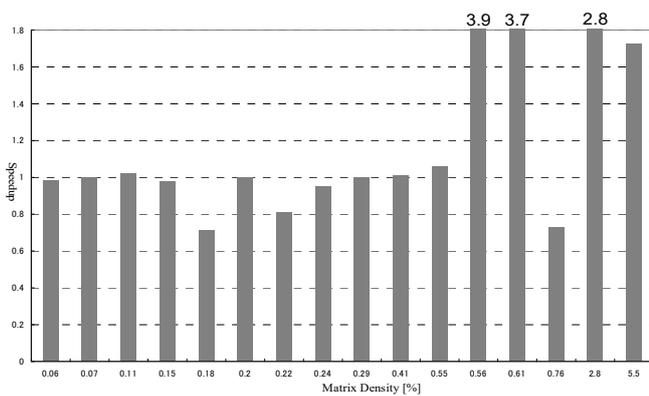

Fig.7 Speedups in SLU' to original SLU.
(Matrix Density: 0.060%--5.5%)

5.4 Performance of RAO-SS
In this section we evaluate the performance of RAO-SS. The aim of this evaluation is two-folded.

First, we evaluate the performance of RAO-SS, compared to the default OP executions in SLU.

Second, we evaluate the software overhead of the Autopilot. This is performed by measuring the execution time between SLU with default OP and RAO-SS in the case of default OP.

Fig.8 and Fig.9 illustrate the execution time with default OP and with RAO-SS. The RAO-SS execution time includes software overhead of the Autopilot.

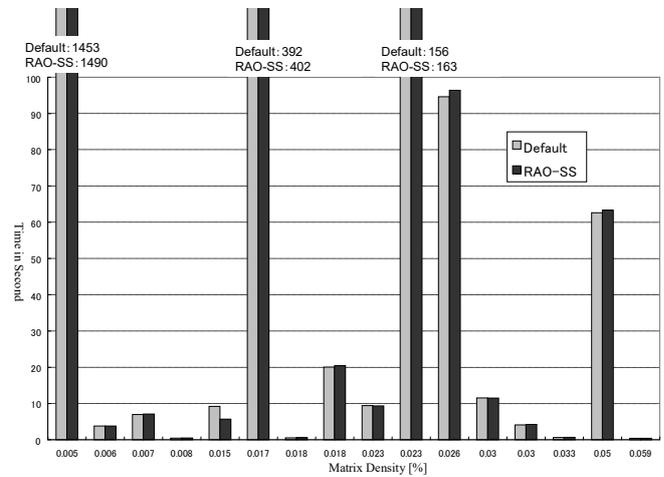

Fig.8 Execution time in RAO-SS and SLU with default OP.
(Matrix Density: 0.005%--0.059%)

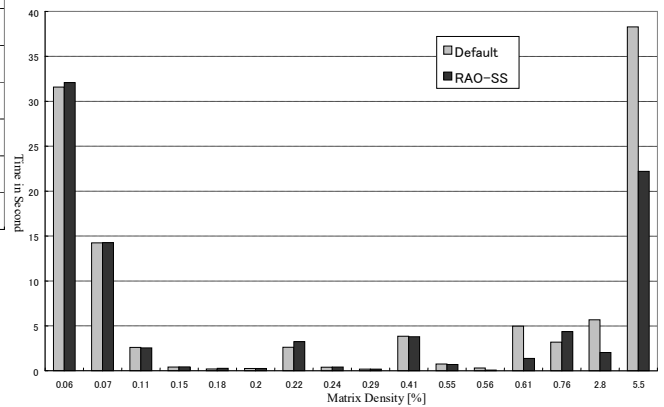

Fig.9 Execution time in RAO-SS and SLU with default OP.
(Matrix Density: 0.060%--5.5%)

The result of Fig.8 and Fig.9 indicated that there were some effects for auto-tuning in RAO-SS. On the other hand, there was no effect for auto-tuning in RAO-SS in some matrices. In this case, the default ordering parameter was selected in RAO-SS. The important point of this case is that the gap between SLU execution time and RAO-SS execution time, since the execution of RAO-SS includes software overhead of the Autopilot.

5.5 Discussion

**For The Best Parameter of SuperLU**

We found the maximum speedup factor of more than 4000 times to the execution with default OP in SLU. The matrix was for the electric circuit problem. We also obtained about 6 times speedup to default OP for the other electric circuit matrix. From this result, the matrices from the electric circuit problems may damage the performance in SLU for default OP execution.

For SLU', we obtained the speedup factors of less than 1.0. Most of these results were from very small execution time. We think, hence, the result was caused by the accuracy of timer function. Of course, the software overhead for the proposed strategy is considerable factor.

**For Software Overhead of RAO-SS**

We obtained the almost same speedup between execution with default OP in SLU and RAO-SS in which the case of selected in default OP. From the result, the software overhead of the Autopilot was negligible in this case. However, we found some speed-down cases, that is, the case of less than 1.0. This causes also the software overhead, but some of them were very small execution time. Hence again, as a conclusion, the overhead of the Autopilot will not be a problem in actual use in such computer environment.

6. Conclusion and Future Work

In this paper, we proposed a facility for auto-tuning for sparse solver, and evaluated the RAO-SS, which is a prototype of run-time auto-tuning software with the Autopilot middle-ware.

With respect to the default parameter execution in the SuperLU for ordering parameters, we obtained the 1.2 times speedup in average, and 3.6 times speedup for maximum with the 32 kinds of test matrices from the University of Florida Sparse Matrix Collection.

As a result of performance evaluation, we conclude that the software overhead of RAO-SS is negligible because the execution time between the SuperLU implemented the proposed strategy and RAO-SS was almost same. This implies the software overhead of the Autopilot was also negligible in the computer environment used in the performance evaluation.

Future work is listed as follows.

- **Developing More Effective Selection Strategy:** Although we found a surprising case that obtains more 4000 times speedup with the best parameter in the SuperLU, we cannot select the best parameter in the proposed strategy. Hence, we need more effective strategy. Determining "effectiveness" for auto-tuning to input matrix sequences, however, is an open problem
- **Development of Dynamically Selection Strategy Function:** If we can define the selection strategy dynamically according to the behavior of input matrices, better parameters may be selected. And we only use the matrix density to estimate matrix characteristics in this experiment. It is not enough measurement to show the characteristics of input matrix. Addition of the length from diagonal elements for non-zero elements[9] may work well against the matrix density. The evaluation is future work. In addition, much use of ability for the Fuzzy Function in the Autopilot, such as dynamically data training function, is important future work to establish dynamic definition for the selection strategy.

・ **Evaluation on the GRID Portal and Heterogeneous Environment:**
The Autopilot runs on the Globus. Thus, RAO-SS also has an ability to run on the GRID environment, namely heterogeneous computer environments. To evaluate RAO-SS in the GRID or heterogeneous computer environments will be a candidate of important roll for auto-tuning software.

Acknowledgements
This study was partially supported by Japan Science and Technology Agency, PRESTO, "Information Infrastructure and Application."

Appendix

Table A-1
Information for the test matrices in University of Florida Sparse Matrix Collection

| No. | Name | Matrix Density |
|---|---|---|
| 1. | *circuit4* | 0.004 |
| 2. | *shyy161* | 0.005 |
| 3. | *epb3* | 0.006 |
| 4. | *hayer01* | 0.008 |
| 5. | *jan99jac120sc* | 0.015 |
| 6. | *mark3jac080sc* | 0.017 |
| 7. | *onetone2* | 0.0175 |
| 8. | *jan99jac100sc* | 0.018 |
| 9. | *jan99jac080sec* | 0.022 |
| 10. | *mark3jac060sc* | 0.023 |
| 11. | *wang4* | 0.026 |
| 12. | *onetone1* | 0.03 |
| 13. | *jan99jac060sc* | 0.03 |
| 14. | *circuit3* | 0.03 |
| 15. | *g7jac080sc* | 0.05 |
| 16. | *ex35* | 0.058 |
| 17. | *g7jac060sc* | 0.06 |
| 18. | *g7jac050sc* | 0.07 |
| 19. | *nmos3* | 0.11 |
| 20. | *lhr14c* | 0.15 |
| 21. | *ex19* | 0.18 |
| 22. | *lhr10c* | 0.2 |
| 23. | *coater2* | 0.22 |
| 24. | *utm5940* | 0.24 |
| 25. | *lhr07c* | 0.29 |
| 26. | *graham1* | 0.41 |
| 27. | *igbt3* | 0.55 |
| 28. | *thermal* | 0.557 |
| 29. | *goodwin* | 0.61 |
| 30. | *ex40* | 0.76 |
| 31. | *raefsky1* | 2.8 |
| 32. | *psmigr_3* | 5.5 |